\documentclass[twocolumn,aps,showpacs]{revtex4}
\usepackage{axodraw}
\usepackage{epsfig}
\newcommand{\slsh}[1]{{\not \! #1}}
\newcommand{\be}{\begin{equation}}
\newcommand{\ee}{\end{equation}}
\newcommand{\bea}{\begin{eqnarray}}
\newcommand{\eea}{\end{eqnarray}}
\newcommand{\M}{{\cal M}}
\newcommand{\nn}{\nonumber}
\begin{document}
\title{Dynamical Fermion Masses and Constraints of Gauge Invariance in
Quenched QED3}
\author{A. Bashir$^\dagger$, A. Raya$^\ddagger$}
\affiliation{$^\dagger$Instituto de F{\'\i}sica y Matem\'aticas,
Universidad Michoacana de San Nicol\'as de Hidalgo, Apartado Postal
2-82, Morelia, Michoac\'an 58040, M\'exico.\\
$^\ddagger$Facultad de Ciencias, Universidad de Colima,
Bernal D\'{\i}az del Castillo \#340, Col. Villa San Sebasti\'an, Colima,
Colima 28045, M\'exico.}

\begin{abstract}

  Numerical study of the Schwinger-Dyson equation (SDE) for the fermion 
propagator (FP)
to obtain dynamically generated chirally asymmetric solution in an arbitrary 
covariant 
gauge $\xi$ is a complicated exercise specially if one employs a 
sophisticated form of the fermion-boson interaction complying with the key 
features of a gauge field theory. However, constraints of gauge invariance 
can help construct such a solution without having the need to solve the
Schwinger-Dyson equation for every value of  $\xi$.
In this article, we propose and implement a method to carry out this task
in quenched quantum electrodynamics in a plane (QED3). We start from an 
approximate analytical 
form of the solution of the SDE for the FP in the Landau gauge. We consider 
the cases in which the 
interaction vertex (i) is bare and (ii) is full.
We then apply the Landau-Khalatnikov-Fradkin 
transformations (LKFT) on the dynamically generated solution and find 
analytical results for arbitrary value of  $\xi$. We also compare our results 
with exact numerical solutions 
available for a small number of values of  $\xi$ obtained 
through a direct analysis of the corresponding SDE.
\end{abstract} 


\maketitle

\section{Introduction}

Quantum electrodynamics in a plane (QED3) continues to attract attention
both in the field of super-conductivity, {e.g.},~\cite{supercon}, where it has 
been used in the study of high $T_c$ super-conductors, as well as in the 
realm of dynamical generation of fundamental fermion masses
where the numerical findings 
on the lattice and the results obtained by employing Schwinger-Dyson equations 
(SDE),~\cite{differences}, are yet to arrive at a final consensus. 

In this  paper, we take up the latter question in the light of SDE.
Owing to its simplicity as compared with
its four-dimensional counterpart and quantum chromodynamics
(QCD), the quenched version of QED3 is particularly neat to unfold the 
intricacies of the SDE for the fermion propagator (FP). 
It provides us with an excellent laboratory to study one of its unwelcome 
features, namely, the lack of gauge invariance of the associated physical 
observables. It hampers a fully reliable predictive power of the 
said equations in the non-perturbative regime of interactions.
This problem can be traced back to not employing, or doing so incorrectly,
the gauge identities such as Ward-Green-Takahashi identities 
(WGTI)~\cite{WGTI}, the 
Nielsen identities (NI)~\cite{Nielsen} and the Landau-Khalatnikov-Fradkin
transformations (LKFT)~\cite{LKF}. In this article, we address this issue 
in the light of the LKFT.

The LKFT of the Green's
functions describe the specific manner
in which these functions transform under a variation of gauge.
These transformations are nonperturbative in nature, 
and they are better described in coordinate space. It is primarily for
this reason that some earlier works on its implementation
in the study of the FP were carried out in the coordinate 
space,~\cite{del}. In the context of gauge technique, the FP in QED4
was found to have correct infrared and ultraviolet behaviour
on employing an ansatz for the longitudinal vertex. Momentum space 
calculations are more tedious, owing to the complications induced by 
Fourier transforms. These difficulties are reflected in ~\cite{bashLKF}
where non perturbative FP is obtained starting from a perturbative
one in the Landau gauge in QED3 and QED4.

In perturbative calculations of the Green's functions, these transformations 
are satisfied at every level of approximation. However, in the 
non-perturbative study of 
gauge field theories through SDE in an arbitrary
covariant gauge, one can ensure these transformations for the FP and the vertex
are satisfied
only with a correct ansatz for the so-called transverse part of the three 
point fermion-boson vertex. In practice it implies constructing a very
complicated vertex if one wants to obtain correct gauge covariance in
all momentum regimes. In QED3, recently an ansatz has been proposed,
~\cite{BR1}
which guarantees that the resulting FP satisfies 
its LKFT to two orders and the vertex itself to the first order in their 
respective perturbative expansions. 
As LKFT are non perturbative in nature, we expect them
not only to be satisfied at every order in perturbation theory but also
in phenomena which are realized only non perturbatively, such as dynamical 
chiral symmetry breaking (DCSB).

%Taking into accout the higher orders 
%would result in complicating
%the fermion-boson interaction so much that solving the SDE for the fermion 
%propagator in its chirally asymmetric form would then be practically 
%impossible.

  The continuum studies of DCSB are carried out through the
SDE for the FP. Gauge dependence of the FP can be obtained 
by solving these equations  in different covariant gauges. 
In 
practice, one needs to go 
in small steps of the gauge parameter away from the Landau gauge and it is 
prohibitively difficult to
be able to compute the result for an arbitrarily large value of the gauge
parameter especially
%This becomes
%a tedious and practically impossible task for a large range of 
%values of $\xi$ 
if a sophisticated form of the three point interaction is
taken into account,~\cite{quenched2,BHR}. 
%However, this difficulty can 
%be avoided
%by solving the said equations only in one particular gauge, where
%computation is not too tedious, and then performing the LKF transform to 
%obtain the result in an arbitrary covariant gauge. 
%We do not expect
%the results obtained in this fashion to be identical in general to those
%stemming from solving the Schwinger-Dyson equation in all gauges.
%We know that the solution of the fermion propagator should vary with gauge 
%in such a fashion that the 
%physical observables associated, such as the condensate, remain gauge 
%invariant. It naturally motivates the need to
%calculate the mass function in other gauges. It is not simple. 
The LKFT provide
us with a mechanism to achieve this goal~\cite{talk}. What we have 
to do is to start
from the result in the Landau gauge and simply perform an LKFT to 
find the result in any other gauge. 

%Moreover, as LKFT are based
%upon gauge symmetry, we stand a better chance in its reflection in the
%calculation of the physical observables.

The paper is organised as follows~:  In Sect.~II we recall the solution
of the FP in the Landau gauge employing the bare vertex.
In order to get better physical insight, we approximate the solution with
a simple analytic parametric form; in Sect.~III, we perform the LKFT of 
the dynamically generated to get the FP in
an arbitrary covariant gauge. In Sect.~IV we discuss in details various 
limiting cases of our results. In Sect.~V, we go further to implement
the full vertex. We present the numerical results in
Sect.~VI and conclusions in Sect.~VII.

\section{The Fermion Propagator for the Bare Vertex}

\begin{figure}[t!] %fig 1
\vspace{-3.5cm}
{\centering
\resizebox*{0.4\textwidth}{0.4\textheight}
{\includegraphics{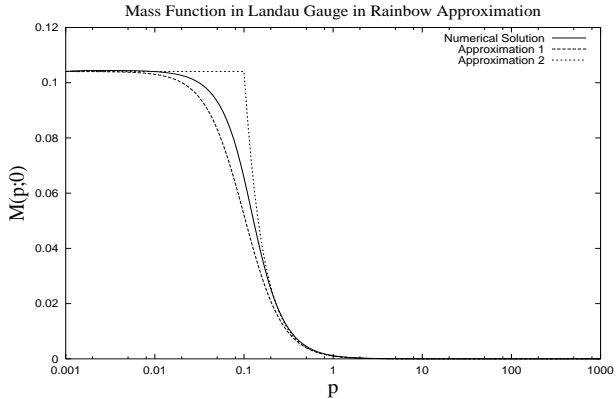}}
\par}
\caption{The mass function in the Landau gauge for the 
bare vertex. Approximations proposed in Eq.~(\ref{param}) and 
Eq.~(\ref{param1}) are also shown.}
\label{fig1}
\end{figure}

%~\footnote{The Diagram
%was generated with AXODRAW~\cite{axodraw}.}
%\begin{center}
%\vspace*{-20pt}
%\begin{picture}(500,100)(0,0)
%Propagador Completo
%\SetScale{0.7}
%\ArrowLine(50,50)(150,50)
%\CCirc(100,50){3}{}{}
%\PText(145,60)(0)[]{-1}
%\PText(100,45)(0)[]{p}
%Propagador desnudo
%\ArrowLine(200,50)(300,50)
%\PText(295,60)(0)[]{-1}
%\PText(250,45)(0)[]{p}
%SDE piece
%\Vertex(430,50){1} \Vertex(370,50){1}
%\ArrowLine(350,50)(450,50)
%\PhotonArc(400,50)(30,0,180){4}{8.5}
%\LongArrowArc(400,50)(20,60,120)
%\PText(400,45)(0)[]{k}
%\PText(400,95)(0)[]{q}
%\CCirc(400,50){3}{}{}
%Signos algebraicos
%\PText(175,52)(0)[]{=}
%\PText(325,52)(0)[]{-}
%\end{picture}\\
%{\sf Diagram~1.  The Fermion Propagator in the Rainbow
%Approximation at One-Loop level.}
%\end{center}
A starting point for the LKF transformations is knowledge of the Green's
functions, here the FP, in the Landau gauge.
We begin this study using the rainbow approximation (where the vertex is
bare) since in this simple case the Landau gauge FP is easy to find.
The well known SDE for the FP in quenched QED3 in the Minkowski 
space is~:
\bea
\hspace{-5mm}
S_M^{-1}(p;\xi)&=& S_M^{0^{-1}}(p) \nn \\
&-&ie^2\int\frac{d^3k}{(2\pi)^3}
\Gamma_M^\nu(k,p)S_M(k;\xi) \gamma^\mu\Delta_{\mu\nu}^{M0}(q),\label{SDEfp}
\eea
where $q=k-p$ and $e$ is the dimensionful electromagnetic
coupling. The bare fermion and photon propagators, respectively, are
$S_M^0(p)=1/{{\not \! p}}$, 
$\Delta_{\mu\nu}^{M0}(q)=-{g_{\mu\nu}}/{q^2}+(1-\xi)
{q_\mu q_\nu}/{q^4}$ and $\xi$ is the covariant gauge parameter.
$\Gamma_M^\mu(k,p)$ is the full fermion-boson vertex which we take
to be bare, i.e., $\gamma^\mu$, for the time being. We shall undo this 
assumption in Sect. V. and implement the full vertex.
We prefer to write the FP in its most general form as
$ S_M(p;\xi)={F(p;\xi)}/({{\not \! p}-{\cal M}(p;\xi)})$. $F(p;\xi)$ 
is referred 
to as the wavefunction renormalization
and $\M (p;\xi)$ as the mass function. The subscript $M$ stands for the 
Minkowski space. Eq.~(\ref{SDEfp}) is a
matrix equation. 
On multiplying it by $1$ and $\slsh{p}$ and taking the
trace, it decomposes into a pair of coupled equations involving $F(p;\xi)$ 
and $\M (p;\xi)$. After a Wick rotation to Euclidean space, and performing 
angular integration, it is a common practice to write these equations as~:
\bea
\frac{1}{F(p;\xi)}&=&1-\frac{\alpha\xi}{\pi p^2} \int dk
\frac{k^2F(k;\xi)}{k^2+{\cal M}^2(k;\xi)}  \nn \\
&& \left[ \;
1-\frac{k^2+p^2}{2kp}\ln{\left|\frac{k+p}{k-p}\right|} \; \right]\;,
\label{eqforF}  \nn \\
\frac{{\cal M}(p;\xi)}{F(p;\xi)}&=&\frac{\alpha(\xi+2)}{\pi p} \int dk
\frac{kF(k;\xi){\cal M}(k;\xi)}{k^2+{\cal M}^2(k;\xi)}
\ln{\left|\frac{k+p}{k-p}\right|} \;, \nn \\ \nn \label{eqforM}
\eea
where $\alpha=e^2/4\pi$. 
 Owing to the fact that in the Landau gauge, $F(p;0)=1$, it has 
long served as a favourite gauge where one only has to solve the 
following equation~:
\be
{\cal M}(p;0)=\frac{2 \alpha }{\pi p}\int_0^\infty dk
\frac{k{\cal M}(k;0)}{k^2+{\cal
M}^2(k;0)}\ln{\left|\frac{k+p}{k-p}\right|}\;.\label{MFint}
\label{LGMF}
\ee
%\begin{center}
%\epsfig{file=rainbowL.ps, width=8.5cm}
%\end{center}
The dynamically generated non perturbative solution of this equation is 
depicted as a solid line in Fig.~(\ref{fig1}). It corresponds to
the appearance of masses through self interactions.
In this article,
we shall apply LKFT on this dynamically generated solution to find the
result in an arbitrary covariant gauge.
As we are interested in the analytical
insight alone for the time being, we try to represent the solution of the 
FP in the  Landau gauge with some simple analytical function.  
We note that for small 
momenta $p$, the solution for the mass function is roughly a constant.
The large-$p$ behaviour of the mass function can be studied by
transforming Eq.~(\ref{MFint}) into a differential equation with
the appropriate boundary conditions~\cite{review}.  Note that
in the integrand we can use the approximation $
\ln{\left|{(k+p)}/{(k-p)}\right|}=({2p}/{k})\theta (k-p)
+({2k}/{p})\theta(p-k)$,
which is valid for $k>>p$ and also for $k<<p$.  Substituting this
expression into Eq.~(\ref{MFint}), we can re-write it in
differential form as (we have rescaled the integral so
that the corresponding solution of the differential equation can
be expressed in units of $e^4$.)~:
\be
    \frac{d}{dp}\left[ p^3 \frac{d {\cal M}(p;0)}{dp} \right] + 
\frac{2}{\pi^2} {\cal M}(p;0)  = 0 \;,
\label{diflog}
\ee
with boundary conditions $p^3 {d \M(p;0)}/{dp}=0|_{p=0}$ and  
$\M(p;0)=0|_{p \rightarrow \infty}$.
The solution of this equation is
\bea
{\cal M}(p;0) = \frac{4}{\pi^2 p} \Bigg[ 
c_1 J_2\left( \sqrt{\frac{8}{\pi^2 p}}\ \right) - 
c_2 Y_2\left( \sqrt{\frac{8}{\pi^2
    p}}\ \right) \Bigg] \;,
\eea 
where $J(x)$ and $Y(x)$ are the Bessel functions of the first and
second kind, respectively. The second boundary condition imposes
$c_2=0$. As $J(1/\sqrt{p}) \rightarrow 1/p$ when $p \rightarrow \infty$,
we conclude that $\M(p;0) \rightarrow 1/p^2$ in this limit. The same
conclusion is reached from the exact result, 
Fig.~(\ref{fig1}), by noticing that in the numerical data,
$p^2 \M(p;0)$ is practically constant near the upper end of the 
momentum region under consideration, namely, $10^{3}$.
Therefore, we approximate 
the numerical solution by the following
function~: 
\be
{\cal M}(p;0)=\frac{M_0 \, m_0^2}{p^2+m_0^2}\;.\label{param}
\ee

%%%%%%%%%%%%%%%%%%%%%%%%%%%%%%%%%%%%%%%%%%%%%%%%
%%%%%%%%%%%%%%%%%%%%%%%%%%%%%%%%%%%%%%%%%%%%%%%%

\noindent
The numerical result suggests us to choose $M_0=0.10406$ and $m_0=0.1$ in 
order that the infrared and ultraviolet
behaviour of the mass function are correctly reproduced.
With this approximation, we find that the resulting Fourier transforms 
involved 
in LKFT (as we shall see later) are still not trivial to perform. Therefore, 
we need a further
simplification. Owing to the fact that the mass function has the following 
infrared and 
ultraviolet features~:
\begin{itemize}
\item region I~: $\; p<<m_0$
\be
\M_{\rm I}(p;0)=M_0\;,       \label{approx1}
\ee

\item region II~: $\; p>>m_0$
\be
\M_{\rm II}(p;0)=\frac{M_0 \, m_0^2}{p^2}\;,  \label{approx2}
\ee
\end{itemize}
we can re-write the approximation to the mass function as follows~:
\be
{\cal M}(p;0)= M_0 \left[ \theta(m_0-p) + \frac{m_0^2}{p^2} \theta(p-m_0) 
\right]
\;. \label{param1}
\ee
As shown in Fig.~(\ref{fig1}), Eqs.~(\ref{param},\ref{param1}) provide a good 
approximation to the mass function in the infrared and ultraviolet
regions. We are now in a position to apply the
LKFT to this approximate solution
to find the corresponding result in an arbitrary covariant gauge. We carry 
out this exercise in the next section.

\section{LKFT and the Dynamically Generated Mass}

  We start by 
putting forward the definitions and notations we shall use along the way.
We write out the FP in momentum and coordinate spaces, 
respectively, in its most general form as~:
\bea
S_E(p;\xi)&=&A(p;\xi)+i\frac{B(p;\xi)}{{\not \! p}}\equiv
\frac{F(p;\xi)}{i {\not \! p}-{\cal M}(p;\xi)}
\;,
\label{fpropmoment} \\
\nonumber   \\ S_E(x;\xi)&=&{\not \! x}X(x;\xi)+ Y(x;\xi)
\label{fpropcoord} \;,
\eea
The subscript $E$ stands for the Euclidean space. The above 
expressions are related through the following Fourier transformations
\bea
S_E(p;\xi)&=&\int d^3x e^{i p\cdot x} S_E(x;\xi)\;, \label{fourierp}\\
S_E(x;\xi)&=&\int \frac{d^3p}{(2\pi)^3} e^{-ip\cdot
x}S_E(p;\xi)  \;.   \label{fourierx}
\eea
The LKFT relating the coordinate space FP in 
Landau gauge to one in an arbitrary covariant gauge reads
\be
S_E(x;\xi)=S_E(x;0)e^{-a x} \;, \label{LKFTp}
\ee
where $a=\alpha \xi/2$. The way we proceed is as follows~:  We start with 
a dynamically generated 
solution of the SDE for the FP in the
Landau gauge and Fourier
transform it to coordinate space. We then apply the LKFT
law. Fourier transform of this result back to the momentum space
yields the FP in an arbitrary covariant gauge.

\subsection{Fermion Propagator in an Arbitrary Gauge in the Coordinate Space}

As we have seen in section II that in the Landau gauge in the rainbow 
approximation, the function $F(p;0)=1$. Therefore, the FP can 
be written $S_E(p;0)= {1}/{(i \slsh{p}-{\cal M}(p;0))}$.
Making use of this expression in Eqs.~(\ref{fpropcoord},\ref{fourierx}),
the coordinate space FP in the Landau gauge can be written as~:
\bea
X(x;0)&=& \frac{1}{2 \pi^2 x^2}  \int_0^{\infty} 
\frac{dp \; p^2}{p^2+{\cal M}^2(p;0)} 
\left[ \cos{px}-\frac{\sin{px}}{px} \right], \nn \\ \label{PxX0}\\
Y(x;0)&=& -\frac{1}{2 \pi^2}  \int_0^{\infty}  
\frac{ dp \; p^2 {\cal M}(p;0)}{p^2+{\cal M}^2(p;0)}
\left[ \frac{\sin{px}}{px} \right]  \;.  \label{PxY0}
\eea
For ${\cal M}(p;0)$, we shall use the approximate solution given in
Eq.~(\ref{param1}).  It naturally divides the integration region in 
Eqs.~(\ref{PxX0},\ref{PxY0}) into two parts. Therefore,
\bea
X(x;0)=X_I+X_{II} \hspace{5mm}{\rm {and}} \hspace{5mm}
Y(x;0)=Y_I+Y_{II}  \;, \label{FP0}
\eea
where
\bea
\hspace{-4mm}
X_I &=& \frac{1}{2\pi^2x^2}\int_0^{m_0}   \frac{dp \; p^2}{p^2+M_0^2} \left[ 
\cos{px}-\frac{\sin{px}}{px}\right] ,  \label{x1} \\
\hspace{-4mm}
X_{II} &=& \frac{1}{2\pi^2x^2}\int_{m_0}^\infty 
\frac{dp \; p^6}{p^6+M_0^2m_0^4} 
\left[\cos{px}-\frac{\sin{px}}{px}\right] , \label{x2} \\
\hspace{-4mm}
Y_I&=&  -\frac{1}{2\pi^2}\int_0^{m_0}  \frac{dp \; p^2 \, M_0}{p^2+M_0^2}
\left[ \frac{\sin{px}}{px} \right] , \label{y1}  \\
\hspace{-4mm}
Y_{II} &=& - \frac{1}{2\pi^2}\int_{m_0}^\infty  
\frac{dp \; p^4 \, M_0 \, m_0^2 }{p^6+ M_0^2 m_0^4} 
\left[ \frac{\sin{px}}{px}\right] .  \label{y2}
\eea
In order to be consistent with the approximations, 
Eqs.~(\ref{approx1},\ref{approx2}) made to arrive at 
Eq.~(\ref{param1}), we neglect the corresponding terms in the denominators
of the Eqs.~(\ref{x1}-\ref{y2}),
depending upon the region of integration over the momentum space. Thus
carrying out the radial integrations, we obtain
\bea
X_I&=&\frac{1}{2\pi^2M_0^2x^5}[ 3m_0x\cos{m_0x} 
+ (m_0^2x^2-3)\sin{m_0x} ]\;, \nn \label{XIxi0}\\
X_{II}&=&-\frac{1}{4\pi^2x^3}\left[ \pi+2\sin{m_0x}-2Si(m_0x)\right]\;, \nn \\
\label{XIIxi0}
Y_I&=&\frac{m_0}{2\pi^2 M_0 x^2} \left[ \cos{m_0x} -
\frac{\sin{m_0x}}{m_0x}\right]
\;,\label{YIxi0} \nn \\
Y_{II}&=& -\frac{M_0 m_0}{4\pi^2}\left[ \cos{m_0x}+
\frac{\sin{m_0x}}{m_0x}\right] 
\nn \\
&+& \frac{M_0 m_0^2 x}{4\pi^2}\left[ \frac{\pi}{2}-Si(m_0x)\right]\;,
\label{YIIxi0}  
\eea
where $Si(y)$ is the Sine Integral function, given by~:
\be
Si(y)=\int_0^y\frac{\sin{t}}{t}dt \;.
\ee
Now Eqs.~(\ref{fpropcoord},\ref{LKFTp},\ref{FP0}-\ref{YIIxi0}) 
constitute the FP
in an arbitrary covariant gauge in the coordinate space.
In the next section, we Fourier transform this result back to the momentum
space.

\subsection{Fermion Propagator in an Arbitrary Gauge in the Momentum Space}

In order to carry out the inverse Fourier transform, we use the auxiliary
functions  $A(p;\xi)$ and $B(p;\xi)$ which are related to the more
familiar functions ${\cal M}(p;\xi)$ and $F(p;\xi)$ in a simple fashion~:
\bea
{\cal M}(p;\xi)&=& p^2 \frac{A(p;\xi)}{B(p;\xi)}      \;,\\
F(p;\xi)&=& -B(p;\xi)-p^2\frac{A^2(p;\xi)}{B(p;\xi)}  \;.
\eea
In terms of $X(x;\xi)$ and $Y(x;\xi)$, the auxiliary functions can 
be written as follows~:
\bea
&& \hspace{-10mm}
A(p;\xi)=\int d^3x e^{ip\cdot x}Y(x;\xi)\nn\\
&& \hspace{1.5mm} = 
\frac{4\pi}{p}\int_0^\infty dx x\sin{px} Y(x;\xi) \;,\\
&& \hspace{-10mm} B(p;\xi)= \frac{1}{i} \; 
\int d^3x \, p \cdot x \, e^{ip\cdot x} X(x;\xi)\nn\\
&& \hspace{1.5mm} =
4\pi
\int_0^\infty dx x^2\left[\frac{\sin{px}}{px}-\cos{px}\right]
X(x;\xi) \;.  
\eea
 We shall eventually
be interested in the large and small limits of the momentum as compared to
the dynamically generated mass scale $m_0$. However, as the coupling also
has mass dimensions, special care may be needed for the evaluation of the
the above integrals, especially for the $B$-term as we shall shortly see.
We evaluate the $A$-term and the $B$-term separately below.

\subsubsection{The $A$-term}

Making use of Eqs.~(\ref{FP0}-\ref{YIIxi0}), $A(p;\xi)$ can be 
expanded term by term as~:
\bea
A(p;\xi)&=&\frac{2m_0}{\pi M_0 p}A_1(p;\xi)
-\frac{M_0 m_0}{\pi p}A_2(p;\xi)\nn\\
&+&\frac{M_0 m_0^2}{\pi p}
A_3(p;\xi)\;,
\eea
where
\bea
A_1(p;\xi)&=& \int_0^\infty \frac{dx}{x} e^{-ax} \sin{px}
\left[ \cos{m_0x}-\frac{\sin{m_0x}}{m_0x}\right]\nn\\
A_2(p;\xi)&=&  \int_0^\infty dx x e^{-ax} \sin{p x} 
\left[ \cos{m_0x}+\frac{\sin{m_0x}}{m_0x}\right]\nn\\
A_3(p;\xi)&=& \int_0^\infty dx x^2 e^{-ax} \sin{px}
\left[ \frac{\pi}{2}-Si(m_0x)\right] \;.
\eea
The results of these integrations have been presented in the 
appendix. The addition and simplification of these expressions
brings about some cancellations to yield the final simple result
as~:
\bea
A(p;\xi)&=& -\frac{1}{\pi}\left[ \frac{1}{M_0}+\frac{M_0 m_0^2 (3a^2-p^2)}
{(a^2+p^2)^3}\right][T_+ +T_-] \nn\\
&+& \frac{a}{2\pi p} \left[\frac{1}{M_0}-\frac{M_0 m_0^2(a^2-3p^2)}
{(a^2+p^2)^3} \right]L\nn\\
&+& \frac{M_0 m_0}{(a^2+p^2)^2} \left[ -\frac{4a}{\pi} + m_0
\frac{(3a^2-p^2)}{a^2+p^2}\right]    \;,  \label{Afunc}
\eea
where
\bea
T_{\pm} &=&\arctan{\left[ \frac{m_0 \pm p}{a}\right]}  \;,\nn\\
L&=&\ln{\left|\frac{a^2+(m_0+p)^2}{a^2+(m_0-p)^2} \right|} \;. 
\label{def1}
\eea

\subsubsection{The $B$-term}

We carry out a similar procedure to evaluate $B(p,\xi)$. However, the
integrations involved are harder as we shall shortly see. 
Eqs.~(\ref{FP0}-\ref{YIIxi0}) permit us to write
\bea
B(p,\xi)&=&
-\frac{6m_0}{\pi M_0^2}B_1(p;\xi)+\frac{2}{\pi}
\left[  1-\frac{m_0^2}{M_0^2} \right]  B_2(p;\xi)  \nn\\
&+& B_3(p;\xi) -\frac{2}{\pi}B_4(p;\xi)\;,
\eea
%\be
%B(p;\xi)=B_1(p;\xi)+B_2(p;\xi)+B_3(p;\xi)+B_4(p;\xi) \;,
%\ee
where the integrals involved are outlined below~:
\bea
&& \hspace{-10mm}
B_1(p;\xi)= \int_0^\infty \frac{dx}{x^2} e^{-a x}
\left[ \cos{m_0x}-\frac{\sin{m_0x}}{m_0x}\right] \nn\\
&&\hspace{20mm}
\left[ \cos{px}-\frac{\sin{px}}{px}\right] \nn\\
&& \hspace{-10mm}
B_2(p;\xi)= \int_0^\infty \frac{dx}{x} e^{-a x} \sin{m_0x}
\left[ \cos{px}-\frac{\sin{px}}{px} \right] \nn\\
&& \hspace{-10mm}
B_3(p;\xi)= \int_0^\infty \frac{dx}{x} e^{-a x}
\left[ \cos{px}-\frac{\sin{px}}{px} \right] \nn\\
&& \hspace{-10mm}
B_4(p;\xi)=\int_0^\infty \frac{dx}{x} e^{-a x}
\left[ \cos{px}-\frac{\sin{px}}{px} \right] Si(m_0x)\;.
\eea
%$B_1,\, B_2$ and $B_3$ can be straightforwardly obtained from
%the  appendix
Except the last one, the other integrals are less complicated to evaluate. 
Making use
of the same notations as before, these are
listed below~:
\bea
B_1(p;\xi)&=& \frac{1}{6m_0p}\left[(m_0^3+p^3)T_+ -(m_0^3-p^3)T_- \right]\nn\\
&&
+\frac{a}{24m_0p}[4m_0p+(a^2+3m_0^2+3p^3)L]\nn\\
B_2(p;\xi)&=& \frac{m_0}{2p} \left[ T_- - T_+\right]  -\frac{a}{4p}L\nn\\
B_3(p;\xi)&=& -1+\frac{a}{p}\arctan{\frac{p}{a}} \;.
\eea
A word of caution is appropriate here. In order to verify our findings, we
shall always want to make a connection of our results for the large and small
values of momentum to the ones in the Landau gauge. It appears that $B_3$ 
may offer a tricky problem in this context because after taking the small
$p$ limit, we cannot retreat to the Landau gauge which corresponds to the
$a \rightarrow 0$ limit. The term $B_4(p;\xi)$ comes to our rescue and it
becomes evident if treated carefully. It is for this reason that we prefer 
to write it in the form~:
\bea
B_4(p;\xi)&=& \sum_{n=0}^\infty (-1)^n \frac{a^n}{n!} I_n\;,
\eea
where
\bea
I_n&=& \int_0^\infty dx x^{n-1} \, \left[ \cos{px}-\frac{\sin{px}}{px}
\right] 
Si(m_0x)\;.
\eea
This expression is perhaps the most convenient to deduce the FP
for the asymptotic limits of momenta and then also to be able
to make connection with the results of the Landau gauge. One can find
out that when $m_0<p$,
\bea
I_n&=& -\frac{m_0}{p^{1+n}} \times \nn\\
&&\hspace{-5mm}
\Bigg[ \Gamma(1+n)\;_pF_q \left[ \left\{
\frac{1}{2},\frac{n+1}{2},1+\frac{n}{2}\right\}, \left\{ \frac{3}{2},
\frac{3}{2}\right\};\frac{m_0^2}{p^2} \right]\nn\\
&&\hspace{-3mm}
+ \, \Gamma(n)\; _pF_q\left[ \left\{\frac{1}{2},\frac{n+1}{2},\frac{n}{2}
\right\},
\left\{ \frac{3}{2}, \frac{3}{2} \right\};\frac{m_0^2}{p^2} \right]\Bigg]\;,
\eea
and when $m_0>p$
\bea
I_n &=& \frac{\pi}{2p^n}\cos{\left(\frac{n\pi}{2}\right)}[\Gamma(n-1)
+\Gamma(n)] +\frac{\Gamma(n)}{nm_0^n}\sin{\left(\frac{n\pi}{2}\right)}\nn\\
& \times &\hspace{-2mm} 
 \Bigg[
\;- \, _pF_q\left[\left\{\frac{n+1}{2},\frac{n}{2},\frac{n}{2} \right\},
\left\{\frac{1}{2}, 1+\frac{n}{2} \right\};\frac{p^2}{m_0^2} \right]\nn\\
&&\hspace{-2mm}
+ \; _pF_q\left[\left\{\frac{n+1}{2},\frac{n}{2},\frac{n}{2} \right\},
\left\{\frac{3}{2}, 1+\frac{n}{2} \right\};\frac{p^2}{m_0^2} \right] \Bigg]
\;.  \label{Inpsmall}
\eea
Finally, Eqs.~(\ref{fpropmoment},\ref{Afunc}-\ref{Inpsmall}) give the
dynamically generated FP in an arbitrary covariant gauge.

\section{Relevant Limits}

We are now in a position to carry out the analytic analysis
of the results obtained in the previous section.
We analyse dynamically generated FP
in the  asymptotic limits of momenta, i.e., when $p>>m_0$ and $m_0>>p$. In 
order to make a quick comparison possible, we write below the expressions
for $A(p;0)$ and $B(p;0)$~:
\bea
\hspace{-6mm}     A(p;0) &=& - \frac{1}{M_0} \left[ 
\theta(m_0 - p) + \frac{M_0^2 m_0^2}{p^4} \theta(p-m_0) \right], 
\label{ALG} \\
\hspace{-6mm}     B(p;0) &=& -   \left[  \frac{p^2}{M_0^4}
\theta(m_0-p) +  \theta(p-m_0)  \right]. \label{BLG}
\eea
We now proceed to evaluate the asymptotic limits of the LKF transformed
functions.

\subsection{Large-$p$ Behaviour of the Fermion Propagator}

In the large-$p$  limit, one can verify that
\bea
A(p;\xi) &=&  \left[ -m_0^2 M_0 - \frac{4 a m_0 (m_0^2 + 3 M_0^2)}{3 M_0 \pi}
\right] \; \frac{1}{p^4}    \nn   \\
&& \hspace{3mm} + {\cal O}\left(\frac{1}{p^6}\right)            \;,  \\
B(p;\xi) &=& -1 + {\cal O}\left(\frac{1}{p}\right) \;.
\eea
The Landau gauge results, Eqs.~(\ref{ALG},\ref{BLG}), are readily 
reproduced if we substitute $a=0$. Correspondingly, the mass function and 
the wavefunction renormalization acquire the following form~:
\bea
{\cal M}(p;\xi)&=&  
\left[ m_0^2 M_0 + \frac{4 a m_0 (m_0^2 + 3 M_0^2)}{3 M_0 \pi} \right] \;
\frac{1}{p^2}+ {\cal O}\left( \frac{1}{p^3}\right) \nn \\
&\equiv& \frac{C_3(\xi)}{p^2}   +{\cal O}\left( \frac{1}{p^3}\right) \;,\\
F(p;\xi) &=&1+{\cal O}\left( \frac{1}{p}\right) \;.
\eea
Again, obviously, if we set $a=0$, we recover the Landau gauge results.
An important implication of these last equations is that in an arbitrary
covariant gauge, the mass function will continue to fall as $1/p^2$ for its
asymptotically large values. Numerically, such an effect was seen although 
for a very small range of values of the gauge parameter $(\xi=0 - 1)$ 
in~\cite{quenched2}. For 
a much broader range $(\xi=0 - 5)$, the same observation was reported 
in~\cite{BHR}. Here, the analytical incorporation of the LKFT
allows us to generalise these findings to an arbitrarily large value of
the gauge parameter. Moreover, the wavefunction renormalization continues
to approach 1 in the large $p$ limit for arbitrarily large values of the 
gauge parameter. This fact is also in accordance with the numerical 
computations done for a limited range of $\xi$~\cite{quenched2,BHR}.
 
\subsection{Low-$p$ Behaviour of the Fermion Propagator}

An analogous analysis for the low-$p$ regime yields
\bea
A(p;\xi)&=& \Bigg[ \frac{m_0 M_0}{a^4}\left( 3m_0-\frac{4a}{\pi}\right)\nn\\
&&
 + \frac{2am_0}{\pi
(a^2+m_0^2)}\left( \frac{1}{M_0}-\frac{m_0^2M_0}{a^4}\right)\nn\\
&&
-\frac{2}{\pi}\left(\frac{1}{M_0}+\frac{3m_0^2M_0}{a^4}\right)
\arctan{\frac{m_0}{a}} \Bigg] + {\cal O}(p^2)\nn\\
&\equiv& C_1(\xi)+{\cal O}(p^2) \;,     \\  \nn \\
B(p;\xi)&=& \frac{2}{3M_0^2\pi}\left[ \frac{am_0(3a^2+5m_0^2-2M_0^2)}
{(a^2+m_0^2)^2}\right.   \nn \\
&& \hspace{12mm} \left.-3\arctan{\frac{m_0}{a}}\right]p^2
+{\cal O}(p^3)\nn\\
&\equiv& C_2(\xi) p^2+{\cal O}(p^3)\;,
\eea
where a neat cancellation has taken place between $B_3$ and the
first sum in $B_4$ (refer to the last equation of the Appendix).
Taking the limit $ a \rightarrow 0 $, we again recuperate the 
corresponding results in Eqs.~(\ref{ALG},\ref{BLG}). The
mass function and the wavefunction renormalization can now
be calculated to be~:
\bea
{\cal M}(p;\xi)&=& \frac{C_1(\xi)}{C_2(\xi)} + {\cal O}(p^2) \;, \\
F(p;\xi)&=&-\frac{C_1^2(\xi)}{C_2(\xi)}-C_2(\xi) p^2  
+ {\cal O}(p^4)\;.
\eea
Expectedly, 

\begin{itemize}

\item

 The mass function is flat for small values of $p$, and it falls off 
as $1/p^2$ for its large values. 
This behaviour was observed for a very limited number of values
of $\xi$ in~\cite{quenched2,BHR} after a complicated numerical
exercise. Here we  prove it to be the case analytically
{\em in an arbitrary covariant gauge}.

\item

The wavefunction renormalization is also a constant for small values
of $p$. For the large values, it approaches 1.  This fact is not
a feature of $F(p^2)$ just in the neighbourhood of the Landau gauge.
It is true independent of our selection of $\xi$.

\end{itemize}

Thus the $p$-dependence of the dynamically generated FP
for the small and large values of the momentum continues to
have the same qualitative features in an arbitrary covariant gauge
as the ones in the Landau gauge and in its 
neighbourhood,~\cite{quenched2,BHR}.

\section{Full Vertex}

  In the previous sections, we have relied on the assumption that
the bare vertex is the correct approximation in the Landau gauge.
Several works indicate that this is not the case. The latest in a 
series of proposals for the full vertex in QED3 is the one suggested 
in~\cite{BR1}. However, its employment to solve the SDE for the
FP is a formidable task even in the Landau gauge.
In perturbation theory, as indicated in~\cite{delba} (see
Eq. (26) of that article), the transverse vertex in fact vanishes in the 
Landau gauge in
the limit $k^2 >> p^2$ for QED in arbitrary dimensions. Based upon this, 
there are several suggestions of 
using a  full vertex whose transverse part vanishes in the Landau gauge. 
In this article, we concentrate on
all such vertices, e.g.,~\cite{BC,BP1,CP1,quenched3}.
In this case, the numerical behaviour of the wavefunction
renormalization modifies to the one shown in Fig.~(\ref{fig2}).
It is no longer 1 for all momentum regimes. Instead, it behaves like a
constant (different from unity) for low momentum, and tends to
one as $p\to\infty$. Therefore, in addition to using the
approximation in Eq.~(\ref{param1}) for the mass function
(with $M_0$ replaced by $M_{0F}$ and $m_{0}$ by
$m_{0F}$ because the solutions in the Landau gauge for the bare vertex and the
full vertex are not identical), we can
use the following simple form for the wavefunction renormalization
in the Landau gauge, Fig.~(\ref{fig2})~:

%%%%%%%%%%%%%%%%%%%%%%%%%%%%%%%%%%%%%%%%%%%%%%%%
\begin{figure}[t!] %fig 1
\vspace{-3.5cm}
{\centering
\resizebox*{0.4\textwidth}{0.4\textheight}
{\includegraphics{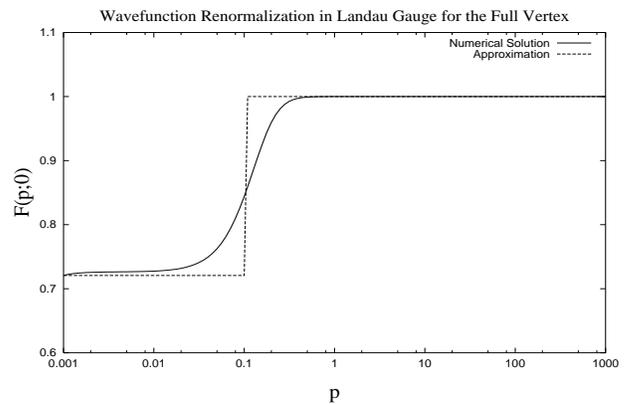}}
\par}
\caption{Wavefunction Renormalization in the Landau gauge for the full
vertex. Approximation proposed in 
Eq.~(\ref{fullparam}) is also shown.}
\label{fig2}
\end{figure}
%%%%%%%%%%%%%%%%%%%%%%%%%%%%%%%%%%%%%%%%%%%%%%%%

\bea
F(p,0) &=& F_{0F} \theta(m_{0F}-p) +\theta(p-m_{0F})  \label{fullparam}  \;.
\eea
Following the steps outlined earlier, it is easy to notice that only the 
terms $X_I$ and $Y_I$ get multiplied by the constant factor $F_0$, whereas,
the rest remains unchanged.  This fact makes the LKFT exercise
straightforward.  The slight modifications to the $A(p;\xi)$ and
$B(p;\xi)$ terms are the following~:
\bea
A(p;\xi)&=& -\frac{1}{\pi}\left[ \frac{F_{0F}}{M_{0F}}+
\frac{M_{0F} m_{0F}^2 (3a^2-p^2)}
{(a^2+p^2)^3}\right][T_+ +T_-] \nn\\
&+&  \frac{a}{2\pi p} \left[\frac{F_{0F}}{M_{0F}}-
\frac{M_{0F}m_{0F}^2(a^2-3p^2)}
{(a^2+p^2)^3} \right]L\nn\\
&+&  \frac{M_{0F} m_{0F}}{(a^2+p^2)^2} \left[ -\frac{4a}{\pi}+ m_{0F}
\frac{(3a^2-p^2)}{a^2+p^2}\right]                     \\
B(p,\xi) &=&
-\frac{6m_{0F}F_{0F}}{\pi M_{0F}^2}B_1+\frac{2}{\pi}
\left[1-\frac{m_{0F}^2F_{0F}}{M_{0F}^2} \right]B_2   \nn \\
 &+& B_3 
-\frac{2}{\pi}B_4 \;, 
\eea
where $B_1,\, B_2,\, B_3$ and $B_4$ have been already calculated.
The large momentum limit for these expressions is now
\bea
A(p;\xi) &=&  \left[ -m_{0F}^2 M_{0F}
 - \frac{4 a m_{0F} (m_{0F}^2 F_{0F} + 3 M_{0F}^2)}{3 M_{0F} \pi}
\right] \; \frac{1}{p^4}    \nn   \\
&& \hspace{3mm} + {\cal O}\left(\frac{1}{p^6}\right)            \;,  \\
B(p;\xi) &=& -1 + {\cal O}\left(\frac{1}{p}\right) \;.
\eea
Consequently,
\bea
{\cal M}(p;\xi)&=&  
\left[ m_{0F}^2 M_{0F} + \frac{4 a m_{0F} (m_{0F}^2 F_{0F} + 
3 M_{0F}^2)}{3 M_{0F} \pi} \right] \;
\frac{1}{p^2} \nn \\
&+& {\cal O}\left( \frac{1}{p^3}\right) \nn \\
&\equiv& \frac{C_{3F}(\xi)}{p^2}   +{\cal O}\left( \frac{1}{p^3}\right) \;,\\
F(p;\xi) &=&1+{\cal O}\left( \frac{1}{p}\right) \;.
\eea
In a similar fashion, for the low-$p$ behaviour, we now arrive at the 
following expressions for $A(p;\xi)$ and $B(p;\xi)$
\bea
A(p;\xi)&=& \Bigg[ \frac{m_{0F} M_{0F}}{a^4}
\left( 3m_{0F}-\frac{4a}{\pi}\right)\nn\\
&&
+ \frac{2am_{0F}}{\pi
(a^2+m_{0F}^2)}\left( \frac{F_{0F}}{M_{0F}}-
\frac{m_{0F}^2M_{0F}}{a^4}\right)\nn\\
&&
-\frac{2}{\pi}\left(\frac{F_{0F}}{M_{0F}}
+\frac{3m_{0F}^2M_{0F}}{a^4}\right)
\arctan{\frac{m_{0F}}{a}} \Bigg]   \nn \\
&& + {\cal O}(p^2)\nn\\
&\equiv& C_{1F}(\xi)+{\cal O}(p^2) \;,     \\  \nn \\
B(p;\xi)&=& \frac{2}{3M_{0F}^2\pi}\left[ 
\frac{am_{0F}(3a^2 F_{0F}+5m_{0F}^2 
F_{0F}-2M_{0F}^2)}
{(a^2+m_{0F}^2)^2}\right.   \nn \\
&& \hspace{12mm} \left.-3 F_{0F} \arctan{\frac{m_{0F}}{a}}\right]p^2
+{\cal O}(p^3)\nn\\
&\equiv& C_{2F}(\xi) p^2+{\cal O}(p^3)\;.
\eea
These functions easily get translated into the more familiar functions
$F(p;\xi)$ and ${\cal M}(p;\xi)$
\bea
{\cal M}(p;\xi)&=& \frac{C_{1F}(\xi)}{C_{2F}(\xi)} + {\cal O}(p^2) \;, \\
F(p;\xi)&=&-\frac{C_{1F}^2(\xi)}{C_{2F}(\xi)}-C_{2F}(\xi) p^2  
+ {\cal O}(p^4)\;.
\eea
We note that the incorporation of the full vertex does not change the 
qualitative features of the mass function and the wavefunction renormalization.
The mass function is flat in the infrared limit and falls off as $1/p^2$
in the ultraviolet region. The wavefunction renormalization is also a constant
for low values of $p$. It approaches 1 as  momentum $p$ acquires large
values. However, the constants multiplying these solutions get modified
on employing the full vertex. To get a quantitative estimate of this
modification we carry out a numerical exercise in the next section.

\section{Numerical Findings}

%%%%%%%%%%%%%%%%%%%%%%%%%%%%%%%%%%%%%%%%%%%%%%%%
\begin{figure}[t!] %fig 1
\vspace{-3.5cm}
{\centering
\resizebox*{0.4\textwidth}{0.4\textheight}
{\includegraphics{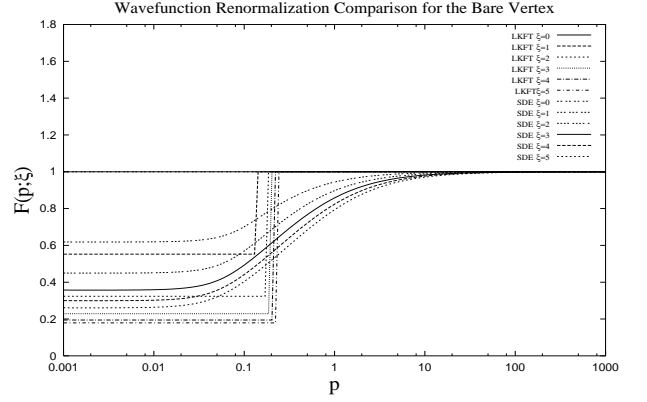}}
\par}
\caption{Wavefunction renormalization for the bare vertex employing LKFT.
For a comparison, we also plot the results obtained by directly solving
SDE.}
\label{fig3}
\end{figure}
%%%%%%%%%%%%%%%%%%%%%%%%%%%%%%%%%%%%%%%%%%%%%%%%

As stressed before, we cannot expect the numerical results presented in this 
section to be quantitatively exact and a true reflection of the choice of
the full vertex owing to the facts that (i) the Landau gauge
result has been approximated by a simple analytic function, and
(ii) approximations~(\ref{approx1},\ref{approx2}) have been employed
in order to ensure that an analytical insight could be obtained. However,
the approximate results indicate that we are making progress in the right 
direction. 

In order to view graphically our numerical results, we proceed as follows.
From the corresponding expressions in the low and large
momentum regimes for $F$ and $\M$, we perform the following
parametrisation for the FP in arbitrary gauge~:
\bea
\M (p;\xi)&=& M_{\xi(F)} \left[ \theta(m_{\xi(F)}-p)
+ \frac{m_{\xi(F)}^2}{p^2} \theta(p-m_{\xi (F)})  \right] \nn \\ \\
F(p;\xi) &=&  F_{\xi(F)} \theta(m_{\xi(F)}-p)
+  \theta(p-m_{\xi (F)})  \;,
\eea
where
\bea
 M_{\xi(F)} &=& \frac{C_{1(F)}(\xi)}{C_{2(F)}(\xi)}  \nn \\
 M_{\xi(F)} m_{\xi(F)}^2 &=& C_{3(F)}(\xi) \nn\\
 F_{\xi(F)} &=& -\frac{C_{1(F)}^2(\xi)}{C_{2(F)}(\xi)} \;. \nn
\eea
and note that it obviously reduces to~(\ref{param1}) 
and~(\ref{fullparam}) on setting $\xi=0$. We separate the discussion
on the bare and the full vertex in the following subsections.

\subsection{Bare Vertex}

%%%%%%%%%%%%%%%%%%%%%%%%%%%%%%%%%%%%%%%%%%%%%%%%
\begin{figure}[t!] %fig 1
\vspace{-3.5cm}
{\centering
\resizebox*{0.4\textwidth}{0.4\textheight}
{\includegraphics{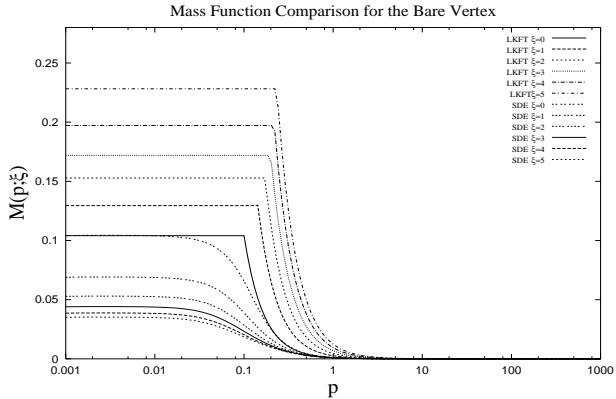}}
\par}
\caption{Mass function for the bare vertex employing LKFT.
For a comparison, we also plot the results obtained by directly solving
SDE.}
\label{fig4}
\end{figure}
%%%%%%%%%%%%%%%%%%%%%%%%%%%%%%%%%%%%%%%%%%%%%%%%

\begin{itemize}

\item

 In Fig.~(\ref{fig3}), we have plotted the wavefunction renormalization
in various gauges. Comparing them with the ones obtained by solving
SDE, one sees that the difference is not enormous, reassuring the 
correctness of the method employed.

\item

 In Fig.~(\ref{fig4}), we have plotted the mass function in several gauges
in order to compare the results of directly solving SDE against the 
ones obtained by employing the LKFT. With the increasing value of the
gauge parameter, the direction of the shift in the mass function 
is diametrically opposed for both the cases. Which one should we trust 
more? LKFT arise as a necessary condition of gauge invariance. Therefore,
one would naturally regard the results more plausible after these 
transformations have been taken into account. We shall see later that the
studies carried out by including the full vertex justify this argument.

\end{itemize}

%%%%%%%%%%%%%%%%%%%%%%%%%%%%%%%%%%%%%%%%%%%%%%%%
\begin{figure}[t!] %fig 1
\vspace{-3.5cm}
{\centering
\resizebox*{0.4\textwidth}{0.4\textheight}
{\includegraphics{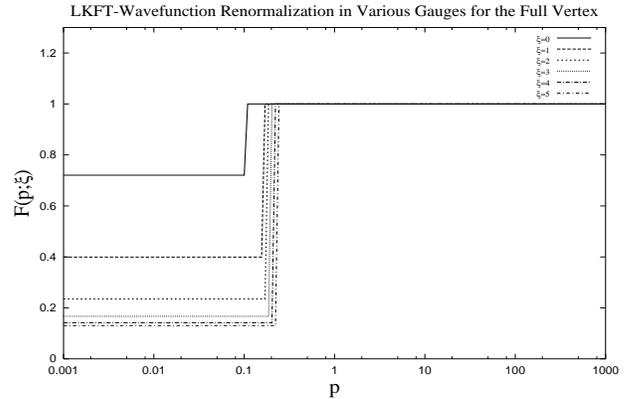}}
\par}
\caption{Wave function renormalization for the full vertex employing LKFT.}
\label{fig5}
\end{figure}
%%%%%%%%%%%%%%%%%%%%%%%%%%%%%%%%%%%%%%%%%%%%%%%%

\subsection{Full Vertex}

SDE for the FP will yield an exact result if a full vertex
is employed which complies with {\em all} the constraints imposed on
it by the corresponding gauge field theory. Construction and usage of
such a vertex is a prohibitively difficult task. Following the works of
Ball and Chiu, we can divide the full vertex into two parts, the longitudinal, 
which ensures WGTI is satisfied and  the transverse which vanishes on 
contraction with the photon momentum vector and remains undetermined by the
WGTI. One of the most recent
(and so far perhaps the most complete) attempts to propose such a vertex in
quenched QED3 can be found in~\cite{BR1}. However, it is obviously a hard 
choice
for its numerical implementation. For the purpose of being able to
compare our results with earlier works we restrict ourselves to the 
family of transverse vertices which vanish in the Landau gauge although
an extension to a more general case is straightforward. We observe the 
following:

%%%%%%%%%%%%%%%%%%%%%%%%%%%%%%%%%%%%%%%%%%%%%%%%
\begin{figure}[t!] %fig 1
\vspace{-3.5cm}
{\centering
\resizebox*{0.4\textwidth}{0.4\textheight}
{\includegraphics{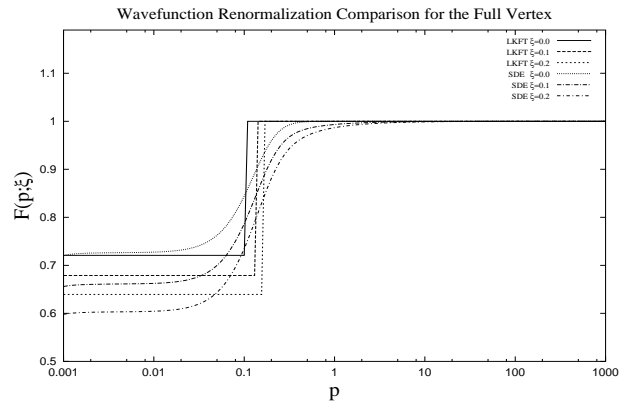}}
\par}
\caption{A comparison of the results for the wavefunction renormalization
(i) employing LKFT and (ii) results obtained by directly 
solving SDE for the full vertex.}
\label{fig6}
\end{figure}
%%%%%%%%%%%%%%%%%%%%%%%%%%%%%%%%%%%%%%%%%%%%%%%%

%%%%%%%%%%%%%%%%%%%%%%%%%%%%%%%%%%%%%%%%%%%%%%%%
\begin{figure}[t!] %fig 1
\vspace{-3.5cm}
{\centering
\resizebox*{0.4\textwidth}{0.4\textheight}
{\includegraphics{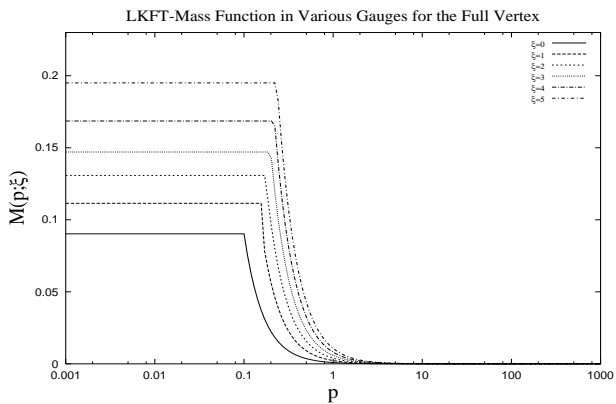}}
\par}
\caption{Mass function for the full vertex employing LKFT.}
\label{fig7}
\end{figure}
%%%%%%%%%%%%%%%%%%%%%%%%%%%%%%%%%%%%%%%%%%%%%%%%

\begin{itemize}

\item In Fig.~(\ref{fig5}), we have plotted the wavefunction renormalization 
in several gauges
and in Fig.~(\ref{fig6}), we present a comparison against  the results of 
directly solving SDEs. The results are found to be in fairly good agreement.

\item In Fig.~(\ref{fig7}), we have plotted the mass function in several 
gauges.
In case of the full vertex, it is a bit hard to compare the results of 
directly solving SDE against the  ones obtained by employing the LKFT
as the results in the former case are known only in a small region near
the Landau gauge~\cite{quenched2,BHR,Huet}. 
We offer a comparison in Fig.~(8). An
interesting thing we note is that in both the cases the trend for
the function is to get shifted upwards with increasing values of the gauge 
parameter. This is the same behaviour as we had noted for the bare vertex 
by using the LKFT. Moreover, the difference between the two types
of approaches seems to close down on each other when the full vertex is 
employed. It is understandable as the constraints of gauge invariance are 
now being used in both the cases. A considerable advantage of using LKFT 
method is that
the results for an arbitrary value of the covariant gauge parameter are
readily available.

\end{itemize}

%%%%%%%%%%%%%%%%%%%%%%%%%%%%%%%%%%%%%%%%%%%%%%%%
\begin{figure}[t!] %fig 1
\vspace{-3.5cm}
{\centering
\resizebox*{0.4\textwidth}{0.4\textheight}
{\includegraphics{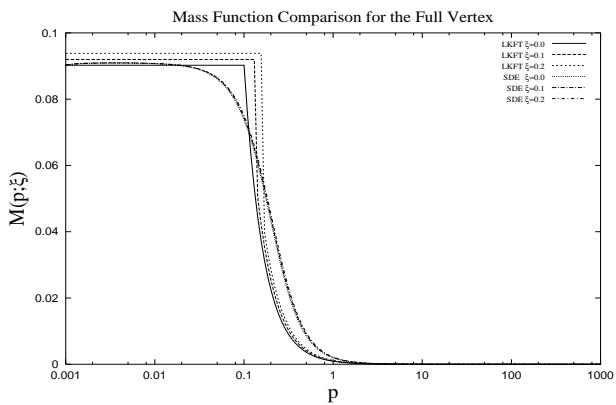}}
\par}
\caption{A comparison of the results for the mass function (i) employing LKFT and (ii) results obtained by directly solving SDE for the full vertex.}
\label{fig8}
\end{figure}
%%%%%%%%%%%%%%%%%%%%%%%%%%%%%%%%%%%%%%%%%%%%%%%%

\section{Conclusions}

The bulk of strong interaction phenomena require a non-perturbative
approach. An example is the dynamical generation of fundamental fermion masses,
which lies beyond the realm of perturbation theory.
This phenomenon is linked with the non-perturbative behaviour
of the FP and is governed by its SDE. As the SDE
of the higher point Green functions are intricately related to
it, a meaningful truncation scheme which maintains the key features of
the theory, such as gauge invariance, continues to be a challenging
problem. Along with other
guiding principles, a correct inclusion of the  LKFT of the FP and the 
fermon-boson interaction is crucial for arriving at reliable conclusions.
We carry out a preliminary study in this connection. Starting from a
dynamically generated chirally asymmetric FP in one gauge,
we perform an LKFT to find its form and properties in an arbitrary covariant 
gauge $\xi$. Encouragingly, we find expected behaviour of the FP for the
asymptotic regimes of momenta for all  $\xi$, a fact supported by earlier
direct numerical solutions of the SDE (where no reference is made
to the LKFT) in a small region of $\xi$ close
to the Landau gauge~\cite{quenched2,BHR} .
The initial results for quenched QED3 presented here provide the starting 
point for a more rigorous and exact numerical study. One can then go
on to take up the more interesting case of QED4. The solution to this problem
will in turn be the staring point for a study of QCD where the non-abelian 
nature of interactions, so essential for both confinement and asymptotic
freedom, will further complicate the problem. All this is for the future.

\vspace*{0.5cm}

\noindent {\bf Aknowledgements:} We whish to thank M. R.
Pennington for suggesting the problem and to him and R. Delbourgo
for their comments on the draft version of this article. 
We acknowledge CIC (project 4.12), CONACyT and Alvarez-Buylla grants.

\vspace*{0.5cm}

\section*{Appendix}

We list below several of the integrals used, \cite{integrals}, 
in this article for a quick reference. Fourier transforming the Landau 
gauge FP in the momentum space to the one in the coordinate
space requires the following integrals~:
\bea
\int_0^m dp p^2 \left[ \cos{px}- \frac{\sin{px}}{px}
\right] &=&
\frac{3mx\cos{mx}}{x^3}\nn\\
&&+\frac{(m^2x^2-3)\sin{mx}}{x^3}\nn \\
\int_m^\infty dp\left[\cos{px}-\frac{\sin{px}}{px}\right] &=&
-\frac{\pi+2\sin{mx}-2Si(mx)}{2x}\nn\\
\int_0^m dp p \sin{p x}&=&
 \left[ -\frac{m\cos{mx}}{x}
+\frac{\sin{mx}}{x^2}\right]\nn\\
\int_m^\infty \frac{dp}{p^3} \sin{px} &=&
\frac{x}{2m}\left[
\frac{\sin{mx}}{mx}+\cos{mx} \right]\nn\\
&&
 -\frac{x^2}{2} \left[
\frac{\pi}{2}-Si(mx) \right] \;.  \nn
\eea
The Fourier trasnsform back to the momentum space after having
performed the LKFT makes use of the following results~:
\bea
A_1(p;\xi) &=& -\frac{p}{2m_0} \left[ T_+ + T_- \right] 
+ \frac{a L}{4m_0}  \nn \\
A_2(p;\xi) &=&  \frac{-p [m_0^4 - (a^2+p^2)^2]}{[a^2+(m_0-p)^2]^2
[a^2+(m_0+p)^2]^2 }    \nn \\ 
A_3(p;\xi)&=& 
-\frac{1}{ (a^2+p^2)^3}\Bigg[ p (3a^2-p^2)(T_+ + T_- -\pi)\nn\\
&& \hspace{-15mm}
+ (a^2-3p^2)L  
+ \frac{4am_0p(a^2+p^2)(2a^2+m_0^2-2p^2)}{A_- A_+}\nn\\
&&
\hspace{-15mm}
+a(a^2+p^2)^2 \Bigg\{  \frac{p(m_0-p)}{A_-^2}
+ \frac{p(m_0+p)}{A_+^2} \nn\\
&&\hspace{-15mm}  + \frac{a^2-(m_0-p)^2}{2A_-^2}
- \frac{a^2-(m_0+p)^2}{2 A_+^2} \Bigg\}  \Bigg]\;, \nn
\eea
where
\bea
A_{\pm} &=& a^2+(m_0 \pm p)^2\;. \nn
\eea
In arrivng at the above relations, we have made use of the following 
relations
\bea
\int_0^\infty dx x^2\sin(px) e^{-a x} &=&
2p \frac{(3a^2-p^2)}{(a^2+p^2)^3}\nn\\
\int_0^\infty \frac{dx}{x} \; e^{-a x}\sin{m_0x} \cos{px} &=&
 \frac{1}{2}  \left[ T_- + T_+ \right] \nn\\
\int_0^\infty dx e^{-a x}\sin{m_0x} \cos{px} &=&
 \frac{m_0(a^2+m_0^2-p^2)}{A_- A_+}\nn\\
\int_0^\infty dx xe^{-a x}\sin{m_0x} \cos{px} &=& 
a\left[ \frac{m_0-p}{A_-^2}+ \frac{m_0+p}{A_+^2}\right]\nn\\
\int_0^\infty \frac{dx}{x}  e^{-a x}\sin{m_0x} \sin{px}&=&
\frac{L}{4} \nn\\
\int_0^\infty dx e^{-a x}\sin{m_0x} \sin{px}&=&
\frac{2am_0p}{A_- A_+}\nn\\
\int_0^\infty dx x e^{-a x}\sin{m_0x} \sin{px}&=&
\frac{1}{2}\Bigg[
\frac{\cos{2T_-}}{A_-}-\frac{\cos{2T_+}}{A_+} \Bigg]\;.\nn
\eea
We have also made use of the following sum
\bea
 \sum_{n=0}^\infty 
\frac{(-a)^n}{n! p^n} \cos{\left(\frac{n\pi}{2}\right)}[\Gamma(n-1)
+\Gamma(n)] = 
-1+\frac{a}{p}\arctan{\frac{p}{a}} .\nn
\eea

%\begin{center}
%\epsfig{file=repmt.ps, width=8.5cm}
%\end{center}

\end{document}